\documentclass[aps,preprint]{revtex4}
\usepackage{graphicx}
\usepackage{epsfig}
\begin{document}
\bibliographystyle{apsrev}
\title{Detecting the Cosmic Gravitational Wave Background with
the Big Bang Observer}
\author{Vincent Corbin and Neil J. Cornish}
\affiliation{Department of Physics, Montana State University, Bozeman, MT
59717}

\begin{abstract}
The detection of the Cosmic Microwave Background Radiation (CMB) was one
of the most important cosmological discoveries of the last century. 
With the development of interferometric gravitational wave detectors,
we may be in a position to detect the gravitational equivalent of the
CMB in this century. The Cosmic Gravitational Background (CGB) is
likely to be isotropic and stochastic, making it difficult to distinguish
from instrument noise. The contribution from the CGB can be isolated by
cross-correlating the signals from two or more independent detectors.
Here we extend previous studies that considered the cross-correlation of two
Michelson channels by calculating the optimal signal to noise ratio that
can be achieved by combining the full set of interferometry variables
that are available with a six link triangular interferometer. In contrast
to the two channel case, we find that the relative orientation of a
pair of coplanar detectors does not affect the signal to noise ratio.
We apply our results to the detector design described in the Big Bang
Observer (BBO) mission concept study and find that BBO could detect
a background with $\Omega_{gw} > 2.2 \times 10^{-17}$.
\end{abstract}

\maketitle

\section{Introduction}

Most theories describing the formation of the Universe (Inflation, Qflation, etc.)
predict that processes in the early Universe will lead to the
copious production of gravitational waves. The detection of such a
cosmic gravitational background (CGB) would allow us to probe the earliest
moments in the history of the Universe, and place strong constraints on
the competing theories~\cite{skc}. However, detecting the CGB will not be easy
since it will be hidden behind the signals from astrophysical sources (binaries
systems...) and buried in the instrumental noise. Our current understanding
of compact binary systems suggests that there is a window above 0.1 Hz where
the number of astrophysical sources is small enough that their contribution can
be isolated and removed from the detector data streams~\cite{curt2,scp}. The CGB signal can then
be dug out of the instrument noise by cross-correlating the outputs of two or
more independent detectors~\cite{flan,allen_romano}. The Big Bang Observer
(BBO)~\cite{BBO_proposal} has been proposed as
a future space based mission designed to operate in the range
$0.1 \rightarrow 1\, {\rm Hz}$. The BBO proposal calls for a fleet of
triangular interferometers operating on the same principle as the Laser
Interferometer Space Antenna (LISA). The BBO detectors will be $\sim 100$
times smaller than the LISA detector, and will be considerably more sensitive.
It is possible to synthesize three independent data channels~\cite{PTLA02} (labeled $(A,E,T)$)
in each detector. In principle one could cross correlate these channels using
data from a single detector as the channels are nominally noise orthogonal.
However, the $A$ and $E$ channels are also signal orthogonal, and the $T$
channel has poor sensitivity to waves with wavelengths larger than the detector.
Moreover, the three channels are constructed from links that share common noise
sources, so it will be difficult to achieve exact noise orthogonality in practice.
The BBO design overcomes these obstacles by employing multiple detector units.
A pair of co-planar detectors yields the greatest sensitivity as the antenna
patterns have significant overlap. The BBO design also calls for two widely
separated ``outrigger'' constellations that give enhanced angular resolution for
detecting astrophysical sources~\cite{Crowder_Cornish}, but the wide separation
($\sqrt{3} {\rm AU} = 866 \; {\rm sec}$) renders them useless for performing
cross correlated detection of the CGB in the $0.1 \rightarrow 1\, {\rm Hz}$
range.

Here we study the optimal cross correlation of co-planar triangular interferometers.
This work generalizes earlier studies~\cite{Cornish_Larson,alberto,njc01} that only considered
the cross correlation of two Michelson channels. By cross-correlating all possible combinations
of the $A,E,T$ channels in the two detectors the overall sensitivity is improved
by a factor of $\sqrt{2}$ at low frequencies and by a factor of $\sqrt{3}$ at high
frequencies. In contrast to the single channel case~\cite{njc01}, the optimal sensitivity is
independent of the relative angle between the two co-planar detectors. We find that
the fiducial BBO design~\cite{BBO_proposal} will permit the detection of a scale invariant
CGB with $\Omega_{gw} = 2.2 \times 10^{-17}$ at 95\% confidence.

We begin in Section II by describing the gravitational wave response of the $A, E$ and
$T$ channels. This is followed in Section III by a calculation of the noise transfer
functions in each channel. Section IV describes the optimal cross correlation of the
independent data channels in a pair of co-planar detectors. Section V covers the
numerical evaluation of the overlap functions and a calculation of the optimal BBO
sensitivity. We use geometric units with $G=c=1$.

\section{Detector Response}

The noise orthogonal data channels $A$, $E$ and $T$ are formed from linear combinations of
the three Sagnac channels $s_{1}, s_{2}, s_{3}$:
\begin{equation}
A=\frac{1}{\sqrt{2}} (s_{3} - s_{1})\qquad E=\frac{1}{\sqrt{6}}
  (s_{1}-2 s_{2}+s_{3}) \qquad T=\frac{1}{\sqrt{3}}
  (s_{1}+s_{2}+s_{3})
\end{equation}
The Sagnac interferometer measures the
phase difference of two laser beams starting from the same
location and going around the triangle formed by the three
spacecraft, one traveling clockwise, the other counterclockwise.
Ideally the phase difference is due only to the variations of the
interferometer arms' length caused by the gravitational waves. Therefore
if the beams start from spacecraft 1, the signal is simply
\begin{equation}
s_{1}(t) = \frac{1}{3 L} [l_{1 3}(t-3 L) + l_{3 2}(t-2 L) + l_{2
1}(t-L) - l_{1 2}(t-3 L) - l_{2 3}(t-2 L) - l_{3 1}(t-L)],
\label{sagnacresponse}
\end{equation}
where $l_{i j}(t - n L)$ is the distance at time $t - n
L$ between spacecraft $i$ and $j$ and $L$ is the length of the
interferometer arms (assuming all the arms have the same length).
For a plane gravitational wave propagating in $\hat{\Omega}$
direction, this can be shown~\cite{njc01} to reduce to
\begin{equation}
s_{1}(t) =
\textbf{D}_{s}(\vec{a}_{1},\vec{b}_{2},\vec{c}_{3},\hat{\Omega},f)
: \textbf{h}(f,t,\vec{x_{1}}),
\end{equation}
where $\vec{a}_{1},\vec{b}_2,\vec{c}_3$ are vectors that point along the interferometer arms,
$\textbf{h}(f,t,\vec{x_{1}})$ is the tensor
describing the wave in the transverse-traceless gauge at point
$\vec{x_{1}}$,
\begin{equation}
\textbf{D}(\vec{a}_{1},\vec{b}_{2},\vec{c}_{3},\hat{\Omega},f) =
\frac{1}{6} \left(\vec{a}\otimes\vec{a} \,\,T_{1}(f,\vec{a}) +
\vec{b}\otimes\vec{b} \,\,T_{2}(f,\vec{b})  +
\vec{c}\otimes\vec{c} \,\,T_{3}(f,\vec{c})\right)
\end{equation}
and
\begin{equation}
T_{1}(\vec{a},f) = e^{-i f_{n} (1+\vec{a} \cdot \hat{\Omega})}
{\rm sinc}\left(f_{n}(1+\vec{a} \cdot \hat{\Omega})\right) - e^{-i f_{n}
(5+\vec{a} \cdot \hat{\Omega})} {\rm sinc}\left(f_{n}(1-\vec{a} \cdot
\hat{\Omega})\right),
\end{equation}
\begin{equation}
T_{2}(\vec{b},f) = e^{-i f_{n} \left[3+(\vec{a}-\vec{c}) \cdot
\hat{\Omega}\right]} \left[ {\rm sinc} \left(f_{n}(1+\vec{b} \cdot
\hat{\Omega})\right) - {\rm sinc}\left(f_{n}(1-\vec{b} \cdot
\hat{\Omega})\right) \right],
\end{equation}
\begin{equation}
T_{3}(\vec{c},f) = e^{-i f_{n} (5-\vec{c} \cdot \hat{\Omega})}
{\rm sinc}\left(f_{n}(1+\vec{c} \cdot \hat{\Omega})\right) - e^{-i f_{n}
(1-\vec{c} \cdot \hat{\Omega})} {\rm sinc}\left(f_{n}(1-\vec{c} \cdot
\hat{\Omega})\right),
\end{equation}
with $f_{n} = \pi L f$. Similarly the Sagnac
signal extracted at vertex 2 and 3 can be found from symmetry by
rotating the system:

\begin{eqnarray}
s_{2}(t) = \textbf{D}_{s}(\vec{c}_{1},\vec{a}_{2},\vec{b}_{3},\hat{\Omega},f) : \textbf{h}(f,t,\vec{x_{2}}), \nonumber \\
\nonumber \\
s_{3}(t) =
\textbf{D}_{s}(\vec{b}_{1},\vec{c}_{2},\vec{a}_{3},\hat{\Omega},f)
: \textbf{h}(f,t,\vec{x_{3}}).
\end{eqnarray}

We are now ready to find the $A$, $E$ and $T$ detector responses to a
plane gravitational wave. To simplify matters we write the signal in the form
\begin{equation}
s_{n}(t) = \textbf{D}_{n}(\hat{\Omega},f) :
\textbf{h}(f,t,\vec{x_{0}}),
\end{equation}
with
\begin{equation}
\textbf{D}_{n}(\hat{\Omega},f) = \vec{a}\otimes\vec{a}
\,\,T_{n}^{a} + \vec{b}\otimes\vec{b} \,\,T_{n}^{b} +
\vec{c}\otimes\vec{c} \,\,T_{n}^{c}.
\end{equation}
We define our reference point $\vec{x_{0}}$ to be the
center of the triangle, and write
\begin{equation}
\textbf{h}(f,t,\vec{x}) = e^{-i 2 \pi f \hat{\Omega} \cdot
(\vec{x}-\vec{x_{0}})} \textbf{h}(f,t,\vec{x_{0}}).
\end{equation}
The $T$ variable, which is also called the symmetrized Sagnac, has an obvious cyclic symmetry:
\begin{equation}
T_{T}^{a} = T_{T}^{b} = T_{T}^{c} = T_{T}, \label{sym}
\end{equation}
where
\begin{eqnarray}
&& T_{T}(\vec{u_{2}} \cdot \hat{\Omega},f)=  \frac{e^{-i \frac{f_n}{3} \bigl(9+(\vec{u_{1}} -
\vec{u_{3}}) \cdot \Omega \bigr)}}{6
\sqrt{3}}  \biggl(1+2 \cos(2 f_n)\biggr)  \biggl[{\rm sinc}\Bigl(f_n(1+\vec{u_{2}} 
\cdot \hat{\Omega})\Bigr) \nonumber \\
&& \hspace*{3.0in}
-{\rm sinc}\Bigl(f_n(1-\vec{u_{2}} \cdot
\hat{\Omega})\Bigr)  \biggr]   \label{TT}
\end{eqnarray}
The variable $A$, on the other hand, does not have this nice
symmetry:
\begin{eqnarray}
&& T_{A}^{a}(\vec{a} \cdot \hat{\Omega},f) = \frac{-i}{3 \sqrt{2}}\sin(f_n) e^{-i \frac{f_n}{3} 
\bigl(6+(\vec{c} - \vec{b}) \cdot \Omega \bigr)} \biggl[{\rm sinc}\Bigl(f_n
(1+\vec{a} \cdot \hat{\Omega})\Bigr) \nonumber \\
&& \hspace*{3.0in}
+e^{-2i f_n} {\rm sinc}\Bigl(f_n(1-\vec{a} \cdot
\hat{\Omega})\Bigr)  \biggr],
\end{eqnarray}

\begin{eqnarray}
&& T_{A}^{b}(\vec{b} \cdot \hat{\Omega},f) = \frac{-i}{3 \sqrt{2}}\sin(f_n)
e^{-i \frac{f_n}{3} \bigl(6+(\vec{a} - \vec{c}) \cdot \Omega \bigr)}
\biggl[ e^{-2i f_n} {\rm sinc}\Bigl(f_n (1+\vec{b} \cdot \hat{\Omega})\Bigr)\nonumber \\
&& \hspace*{3.0in}
+{\rm sinc}\Bigl(f_n(1-\vec{b} \cdot
\hat{\Omega})\Bigr)  \biggr],
\end{eqnarray}

\begin{eqnarray}
&& T_{A}^{c}(\vec{c} \cdot \hat{\Omega},f) = \frac{-i}{3 \sqrt{2}}\sin(f_n) 
e^{-i \frac{f_n}{3} \bigl(9+(\vec{b} - \vec{a}) \cdot \Omega \bigr)} 
\biggl[{\rm sinc}\Bigl(f_n(1+\vec{c} \cdot \hat{\Omega})\Bigr)\nonumber \\
&& \hspace*{3.0in}
+{\rm sinc}\Bigl(f_n(1-\vec{c} \cdot
\hat{\Omega})\Bigr)  \biggr],
\end{eqnarray}

\noindent nor does the variable $E$:
\begin{eqnarray}
&& T_{E}^{a}(\vec{a} \cdot \hat{\Omega},f) = \frac{1}{6 \sqrt{6}}
\biggl[ {\rm sinc}\Bigl(f_n(1+\vec{a} \cdot
\hat{\Omega})\Bigr) \Bigl(e^{-i \frac{f_n}{3} \bigl(9+(\vec{c}
- \vec{b}) \cdot \hat{\Omega} \bigr)} + e^{-i \frac{f_n}{3}
\bigl(3+(\vec{c} - \vec{b}) \cdot \hat{\Omega} \bigr)} \nonumber \\
&& \hspace*{1.0in}
- 2 e^{-i
\frac{f_n}{3} \bigl(15+( \vec{c}
-  \vec{b}) \cdot \hat{\Omega} \bigr)}\Bigr) +{\rm sinc}\Bigl(f_n(1-\vec{a} \cdot \hat{\Omega})\Bigr)
\Bigl( 2 e^{-i \frac{f_n}{3} \bigl(3+(\vec{c} -  \vec{b})
\cdot \hat{\Omega} \bigr)}  \nonumber \\
&& \hspace*{1.0in}
- e^{-i \frac{f_n}{3} \bigl(15+(
\vec{c} - \vec{b}) \cdot \hat{\Omega} \bigr)} - e^{-i \frac{f_n}{3}
\bigl(9+(\vec{c} - \vec{b}) \cdot \hat{\Omega} \bigr)} \Bigr) \biggr],
\end{eqnarray}

\begin{eqnarray}
&& T_{E}^{b}(\vec{b} \cdot \hat{\Omega},f) = \frac{1}{6 \sqrt{6}}
\biggl[ {\rm sinc}\Bigl(f_n(1+\vec{b} \cdot
\hat{\Omega})\Bigr) \Bigl(e^{-i \frac{f_n}{3}
\bigl(15+(\vec{a} - \vec{c}) \cdot \hat{\Omega} \bigr)} + e^{-i
\frac{f_n}{3} \bigl(9+( \vec{a} - \vec{c}) \cdot \hat{\Omega}
\bigr)}  \nonumber \\
&& \hspace*{1.0in}
- 2 e^{-i \frac{f_n}{3} \bigl(3+( \vec{a}
- \vec{c}) \cdot \hat{\Omega} \bigr)}\Bigr)
 +{\rm sinc}\Bigl(f_n (1-\vec{b} \cdot \hat{\Omega})\Bigr)
\Bigl( 2 e^{-i \frac{f_n}{3} \bigl(15+( \vec{a} - \vec{c})
\cdot \hat{\Omega} \bigr)} \nonumber \\
&& \hspace*{1.0in}
- e^{-i \frac{f_n}{3} \bigl(9+( \vec{a} -
\vec{c}) \cdot \hat{\Omega} \bigr)} - e^{-i \frac{f_n}{3}
\bigl(3+(\vec{a} - \vec{c}) \cdot \hat{\Omega} \bigr)} \Bigr) \biggr],
\end{eqnarray}

\begin{eqnarray}
&& T_{E}^{c}(\vec{c} \cdot \hat{\Omega},f) = \frac{1}{6 \sqrt{6}}
\biggl[ {\rm sinc}\Bigl(f_n(1+\vec{c} \cdot
\hat{\Omega})\Bigr) \Bigl(e^{-i \frac{f_n}{3} \bigl(3+(
\vec{b} -  \vec{a}) \cdot \hat{\Omega} \bigr)} + e^{-i \frac{f_n}{3}
\bigl(15 +( \vec{b} - \vec{a}) \cdot \hat{\Omega} \bigr)} \nonumber \\
&& \hspace*{1.0in}
- 2 e^{-i \frac{f_n}{3} \bigl(9+( \vec{b}- \vec{a}) \cdot \hat{\Omega} \bigr)}\Bigr)
 +{\rm sinc}\Bigl(f_n (1-\vec{c} \cdot \hat{\Omega})\Bigr)
\Bigl( 2 e^{-i \frac{f_n}{3} \bigl(9+( \vec{b} - \vec{a})
\cdot \hat{\Omega} \bigr)} \nonumber \\
&& \hspace*{1.0in}
- e^{-i \frac{f_n}{3} \bigl(3+(
\vec{b} - \vec{a}) \cdot \hat{\Omega} \bigr)} - e^{-i \frac{f_n}{3}
\bigl(15+(\vec{b} - \vec{a}) \cdot \hat{\Omega} \bigr)} \Bigr)
\biggr].  \label{TE}
\end{eqnarray}

\section{Noise Spectral Density}

Until now we have only considered the gravitational wave
contribution, $\psi_{ij}(t)$, to the time-varying part of
the phase $\phi_{ij}(t)$. Our next task is to account for
the instrument noise contributions. There are three main
noise sources: the laser phase noise  $C(t)$,  the position noise
$n^{p}(t)$ and the acceleration noise $n^{a}(t)$.

The total phase variation is given by
\begin{equation}
\Phi_{i j}(t) = C_{i}(t - L_{i j}) - C_{j}(t) + \psi_{i j}(t) +
n_{i j}^{p}(t) - \hat{x}_{i j} \cdot [\vec{n}_{i j}^{a}(t) -
\vec{n}_{j i}^{a}(t-L_{i j})].
\end{equation}
The position noise $n^{p}_{ij}(t)$ includes shot noise and pointing jitter
in the measurement of the signal sent by spacecraft $i$ and measured
by the photo-detector in spacecraft $j$. The acceleration noise
$\vec{n}^{a}_{ij}(t)$ is from the gravitation reference system in spacecraft $j$
along the axis that points toward spacecraft $i$. The phase noise associated
with the laser on spacecraft $i$ is denoted $C_{i}(t)$. It is easy to show
that the phase noise cancels in a rigid, non-rotating Sagnac interferometer.
More complicated second generation Sagnac variables can be constructed to
account for the rotation and flexing of the array\cite{neil_ron,dan}. For simplicity we work
with the basic Sagnac variables as they give results that are almost identical
to those found using the second generation variables.

We assume that all the interferometer arms are of approximately equal length
($L= 5 \times 10^{7} \, {\rm m}$), and that the noise spectral densities $S_n(f)$
are similar on each spacecraft. The noise transfer functions are then given by
\begin{equation}
S^{T}_{n}(f) = 2 \left[1 +
2 \cos(2 f_n)\right]^{2} \left[S^{p}_{n}(f) + 4
\sin^{2}(f_n) S^{a}_{n}(f)\right],
\end{equation}
where according to the fiducial BBO design
\begin{eqnarray}
S_{n}^{p}(f) = \frac{2.0 \times 10^{-34}}{(3L)^{2}} \,{\rm Hz}^{-1},  \nonumber \\
S_{n}^{a}(f)  = \frac{9.0
\times 10^{-34}}{(2 \pi f)^{4}(3L)^{2}} \, {\rm Hz}^{-1}.
\end{eqnarray}
This expression for the Symmeterized Sagnac noise transfer function was
previously derived in Refs.~\cite{njc01,PTLA02}. The noise transfer functions
in the $A$ and $E$ channel share the same form, as first pointed out in
Ref.~\cite{PTLA02}:
\begin{eqnarray}
S^{A}_{n}(f) = S^{E}_{n}(f) = 8 \sin^{2}(f_n) \biggl[
\Bigl(2 + \cos(2 f_n) \Bigr)S^{s}_{n}(f) + 2 \Bigl(3 + 2
\cos(2 f_n) + \cos(4 f_n) \Bigr)
S^{a}_{n}(f)\biggr].
\end{eqnarray}

\section{Cross-Correlation of Two Detectors}

The CGB signal can be extracted from the instrument noise by cross-correlating the
outputs of two independent interferometers. The pair of co-planar interferometers
of the BBO do not share any common components, so the noise in each detector should
be largely uncorrelated. Possible correlated sources of noise include solar flares
and fluctuations in the refractive index of the inter-planetary medium. Another potential
source of correlated noise is the residual from subtracting foreground sources
such as double neutron star binaries~\cite{curt2,scp}. Here we will
assume that any correlated sources of noise are well below the level of the CGB.

We assume that the CGB is stationary, Gaussian, isotropic, and unpolarized. The
background can be expanded in terms of plane waves:
\begin{equation}
h_{i j}(t,\vec{x}) = \sum_{A} \int_{-\infty}^{\infty} df \int
d\hat{\Omega} \, \tilde{h}_{A}(f,\hat{\Omega}) e^{-2 \pi i f (t -
\hat{\Omega}\cdot \vec{x})} \epsilon^{A}_{i
j}(\hat{\Omega}),
\end{equation}
where ${\epsilon}^{A}(\hat{\Omega})$ are polarization tensors given in term of the basis tensors
\begin{eqnarray}
\epsilon^{+}(\hat{\Omega},\psi) = \textbf{e}^{+}(\hat{\Omega}) \cos 2\psi - \textbf{e}^{\times}(\hat{\Omega}) \sin 2\psi,
\nonumber \\
\epsilon^{\times}(\hat{\Omega},\psi) = \textbf{e}^{+}(\hat{\Omega}) \sin 2\psi + \textbf{e}^{\times}(\hat{\Omega}) \cos 2\psi.
\end{eqnarray}
The basis tensors can be expressed in term of an orthonormal set of unit vectors $\hat{m}$, $\hat{n}$ and $\hat{\Omega}$ as
\begin{eqnarray}
\textbf{e}^{+}(\hat{\Omega}) = \hat{m}\otimes\hat{m} - \hat{n}\otimes\hat{n},
\nonumber \\
\textbf{e}^{\times}(\hat{\Omega}) = \hat{m}\otimes\hat{n} + \hat{n}\otimes\hat{m}.
\end{eqnarray}
From the strain
\begin{equation}
S(\hat{\Omega},f,t) = \textbf{D}(\hat{\Omega},f) : \bigl(h^{+}(f,t,\vec{x_{0}})\epsilon^{+}(\hat{\Omega},\psi) + h^{\times}(f,t,\vec{x_{0}}) \epsilon^{\times}(\hat{\Omega},\psi)\bigr),
\end{equation}
we find that after averaging over polarizations, the cross-correlated signal
of two detectors in the presence of such a background is given by
\begin{equation}
\langle S_{1}(t) S_{2}(t) \rangle = \int_{0}^{\infty} df S_{h}(f)
R_{1 2}(f),
\end{equation}
where $S_h(f)$ is the total power spectral density due to both polarizations:
\begin{equation}
S_{h}(f) = S^{+}_{h}(f) + S^{\times}_{h}(f),
\end{equation}
and
\begin{equation}
R_{1 2}(f) = \sum_{A} \int \frac{d \Omega}{8 \pi}
{F}_{1}^{A *}(\hat{\Omega},f)
{F}_{2}^{A}(\hat{\Omega},f)  \label{transfer}.
\end{equation}
The transfer function $R_{1 2}$ is a purely geometric factor that accounts for the
overlap of the antenna patterns of the two detectors. The antenna pattern functions
$F$ are given by
\begin{equation}
{F}^{A}(\hat{\Omega},f) = \textbf{D}(\hat{\Omega},f) :
\textbf{e}^{A}(\hat{\Omega}).
\end{equation}

The optimal signal-to-noise ratio squared is given by~\cite{flan,allen_romano}
\begin{equation}
{\rm SNR}_{C}^{2}=2 T \int^{\infty}_{0}df
 S_{h}^{2}(f)\frac{\vert R_{12}(f) \vert^{2}}{S_{n_{1}}(f)
 S_{n_{2}}(f)}.
 \label{SNR1}
 \end{equation}
where $S_{n_{1}}$ is the noise spectral density of
interferometer 1, $S_{n_{2}}$  is the noise spectral density of for interferometer 2
and $S_{h}$ is the spectral density of the CGB.
A signal to noise ratio of
${\rm SNR} = 3.3$ indicates that the CGB has been detected at 95\% confidence, with
a 5\% false alarm probability~\cite{allen_romano}.

The power spectral density of the CGB is related to the energy density in gravitational
waves per logarithmic frequency interval, $\Omega_{gw}(f)$ (in units of the critical
denisty), by
\begin{equation}
S_{h}(f) = \frac{3 H_{0}^{2}}{4 \pi^{2}}
\frac{\Omega_{gw}}{f^{3}}.
\end{equation}
Standard inflationary models predict that $\Omega_{gw}(f)$ will be roughly scale invariant,
with an amplitude $\Omega_{gw} \sim 10^{-15}$ in the $f=1\,{\rm Hz}$ region. 

The contribution to the cross-correlated SNR per logarithmic frequency interval can be written as
\begin{equation}
\frac{ d \, {\rm SNR}^2_C}{ d \ln f} = \frac{h_{\rm opt}^4(f)}{h_{\rm eff}^4(f)}
\end{equation}
where $h_{\rm eff}(f)$ is the effective sensitivity curve
\begin{equation}
\tilde{h}_{\rm eff}(f) = \sqrt{\frac{S_n(f)}{\vert R_{12}(f) \vert}}
\label{effsensitivity}
\end{equation}
and $h_{\rm opt}(f)$ is the optimally filtered CGB signal
\begin{equation}
h_{\rm opt}(f) = (2 T f)^{1/4} \sqrt{S_h(f)}\, .
\end{equation}
These definitions are equivalent to the usual definitions used to plot sensitivity curves
and gravitational wave signals for coherent gravitational wave sources. The main difference is
that the optimally filtered signal strength grows as $T^{1/2}$ with coherent matched filtering,
while it only grows as $T^{1/4}$ for cross-correlated stochastic signals.

\section{Numerical Analysis}

The optimal SNR for the BBO comes from combining the full set of independent interferometry
channels in each of the two co-located detectors that form the star constellation:
\begin{equation}\label{opt_combo}
{\rm SNR}^2_{\rm opt} = \sum_{\alpha = A_1, E_1, T_1}\sum_{\beta = A_2, E_2, T_2} {\rm SNR^2_{\alpha \beta}} \, .
\end{equation}
Our first task is to calculated the geometrical overlap of each pair of channels, $R_{\alpha \beta}(f)$.
The all-sky integral in (\ref{transfer}) was performed numerically using the {\it HEALPIX} package~\cite{healpix}.
Plots of the overlap factors are shown in Figures 1-6 for the standard BBO configuration in which the
two overlapping detectors form a symmetric six pointed star. The plots show the $R$'s scaled by overall
factors such as $\sin^{2}(f_n)$, which they share in common with the corresponding noise transfer functions.

\begin{figure}[ht]
\includegraphics[angle=270,width=0.7\textwidth]{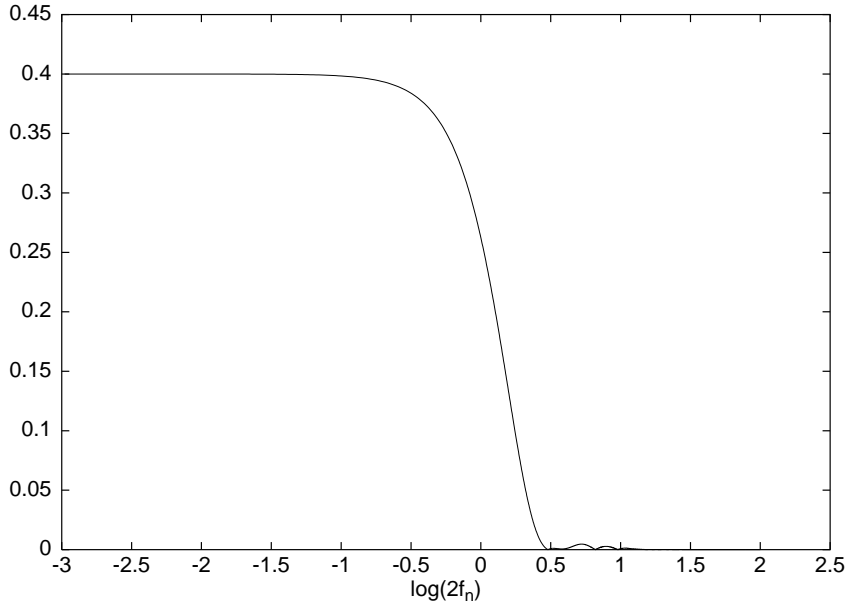}
\caption{$R_{12}(f_{n})/\sin^{2}(f_n)$ for $A_{1} \times A_{2}$}
\end{figure}
\begin{figure}[ht]
\includegraphics[angle=270,width=0.7\textwidth]{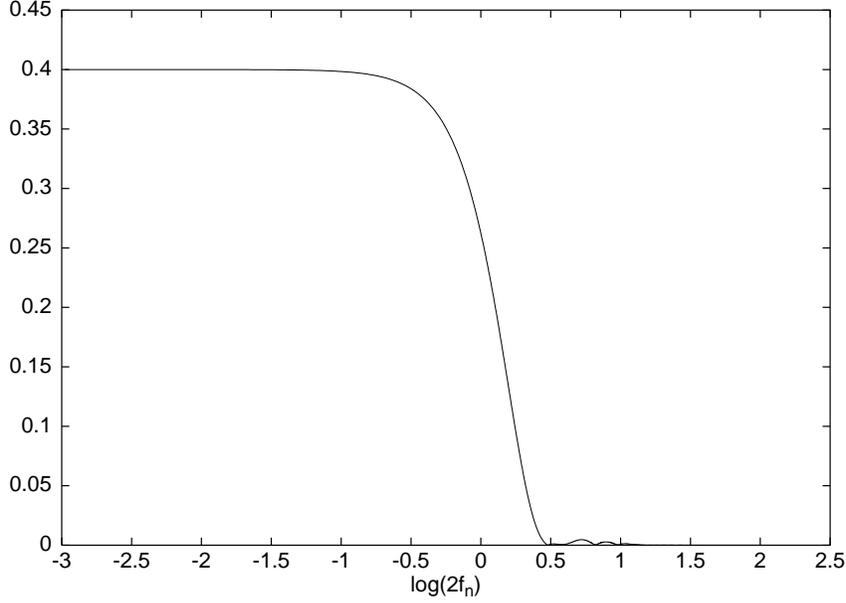}
\caption{$R_{12}(f_{n})/\sin^{2}(f_n)$ for $E_{1} \times E_{2}$}
\end{figure}
\begin{figure}[ht]\label{RTT}
\includegraphics[angle=270,width=0.7\textwidth]{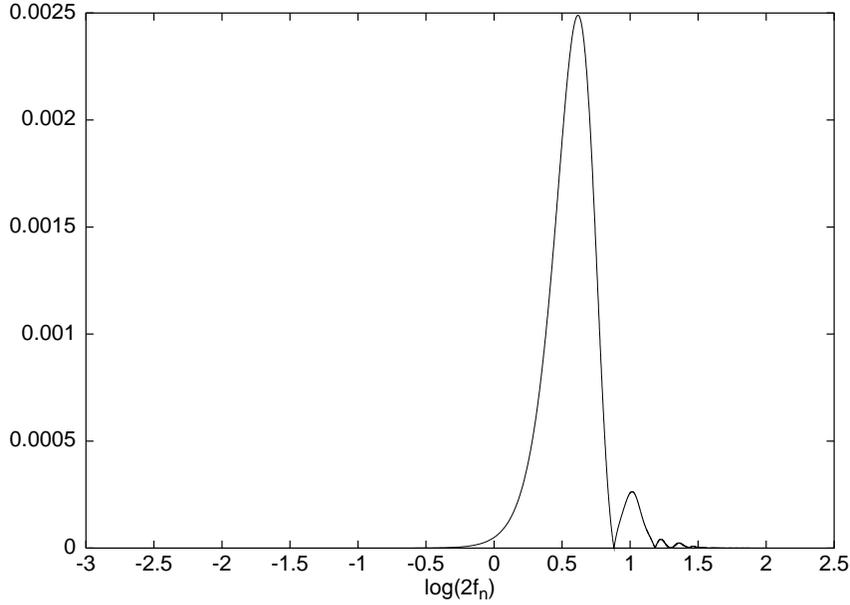}
\caption{$R_{12}(f_{n})/(1+2\cos(2f_n))^{2}$ for $T_{1} \times T_{2}$}
\end{figure}
\begin{figure}[ht]
\includegraphics[angle=270,width=0.7\textwidth]{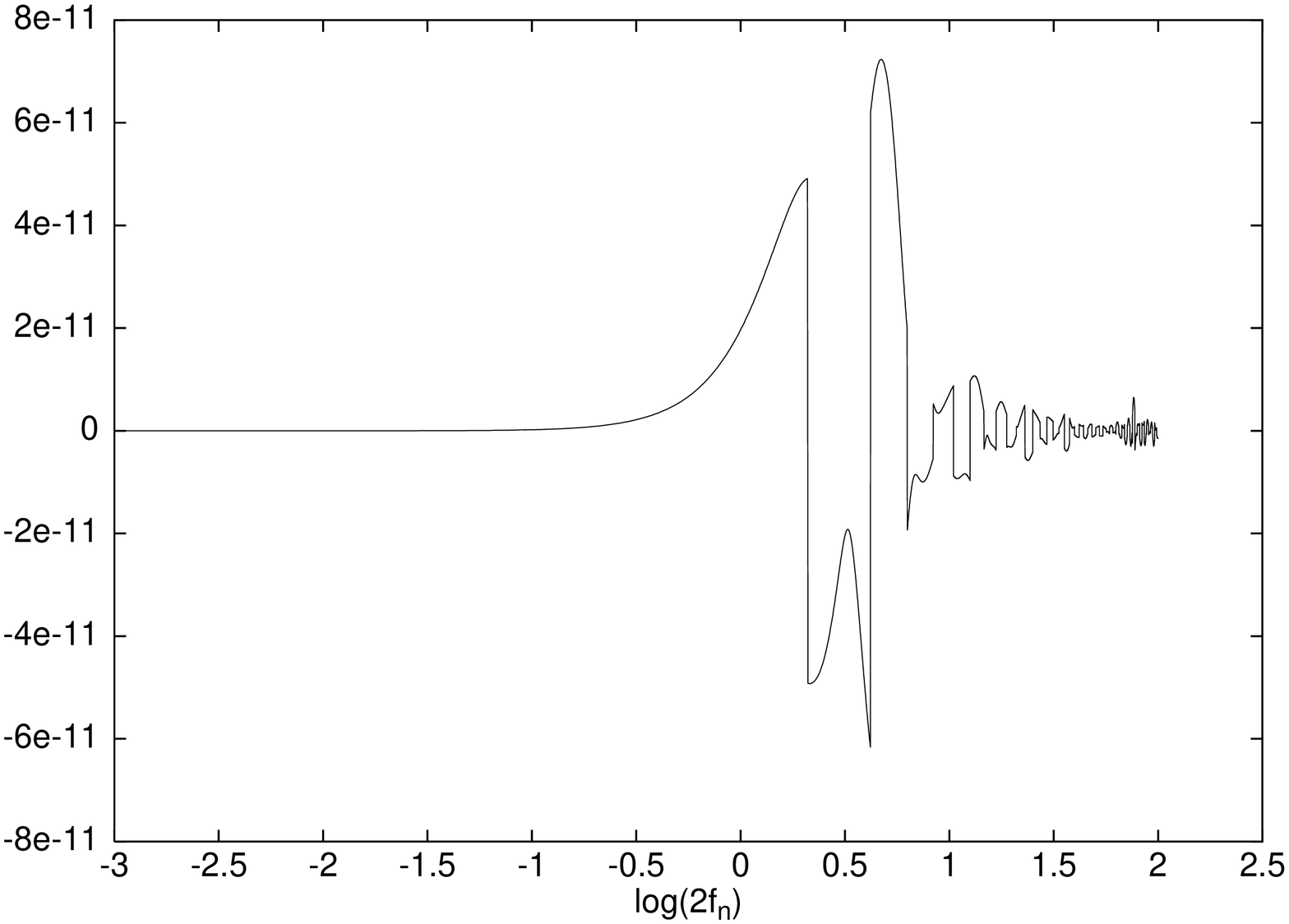}
\caption{$R_{12}(f_{n})/((1+2\cos(2 f_n))\sin(f_n))$
for $A_{1} \times T_{2}$ or $T_{1} \times A_{2}$}
\end{figure}
\begin{figure}[ht]
\includegraphics[angle=270,width=0.7\textwidth]{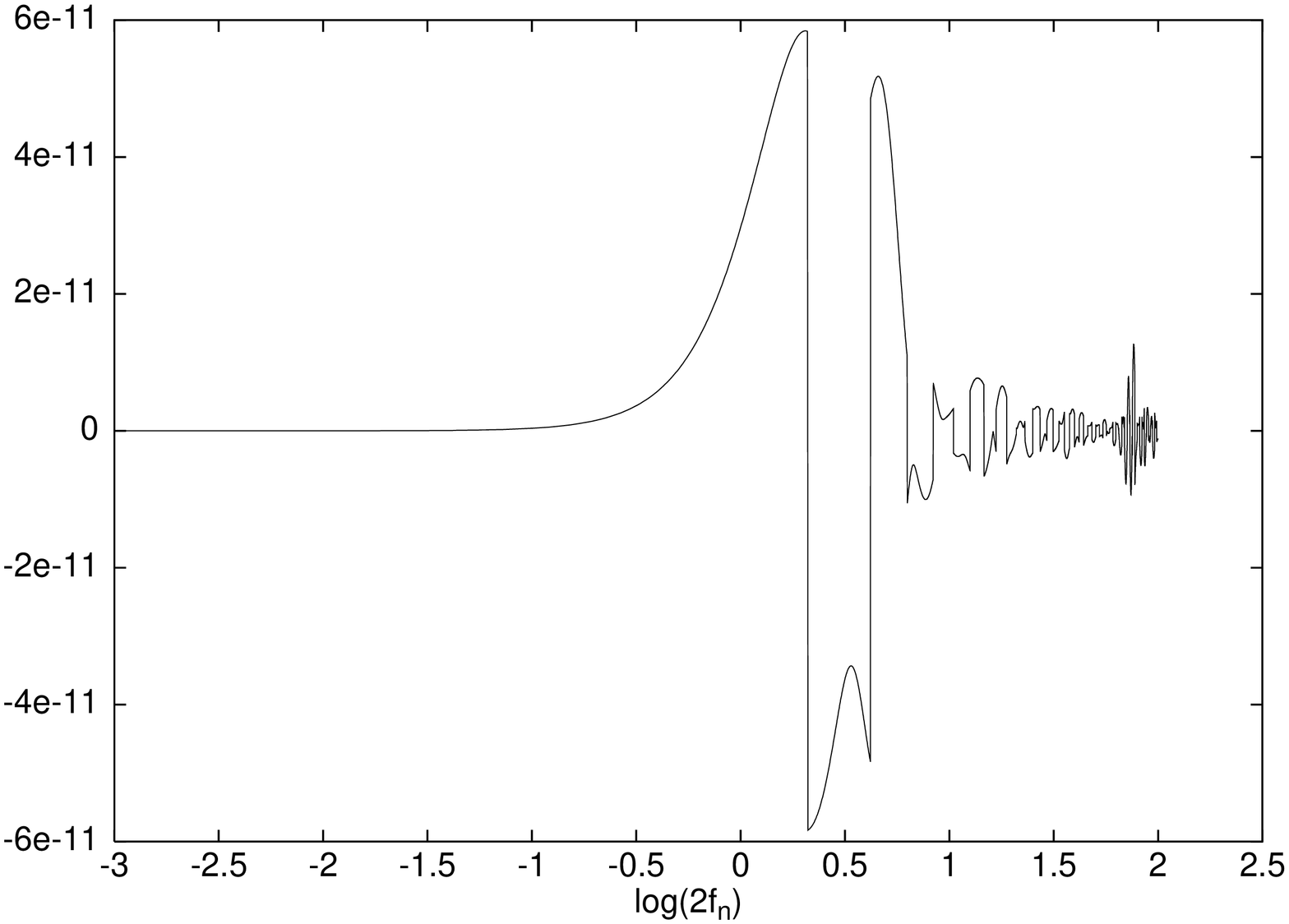}
\caption{$R_{12}(f_{n})/((1+2\cos(2 f_n))\sin(f_n))$
for $E_{1} \times T_{2}$ or $T_{1} \times E_{2}$.}
\end{figure}

\begin{figure}[h]
\includegraphics[angle=270,width=0.7\textwidth]{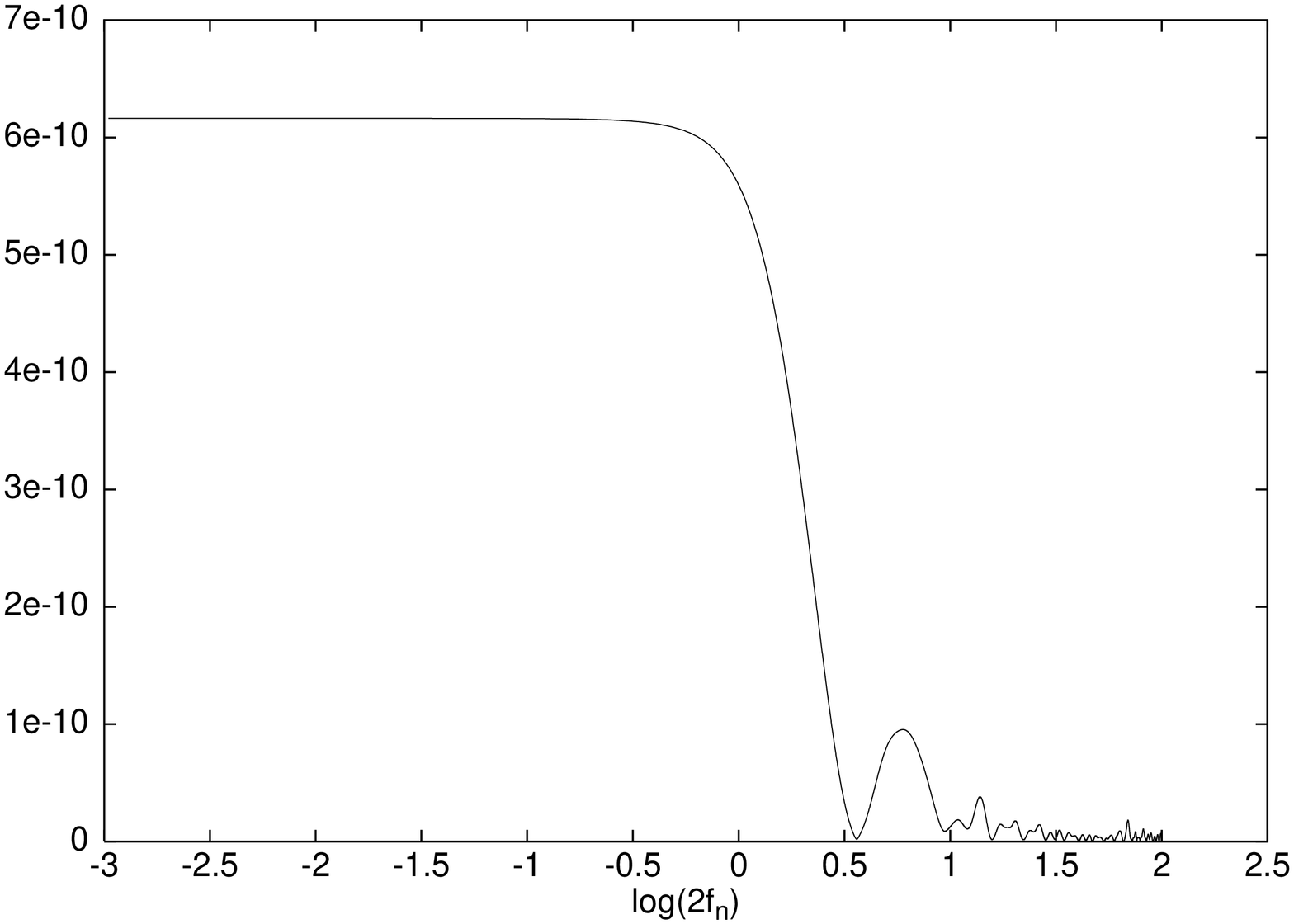}
\caption{$R_{12}(f_{n})/\sin^{2}(f_n)$ for $A_{1} \times E_{2}$
or $E_{1} \times A_{2}$.}
\end{figure}

We expect $R_{12}$ to vanish for the cross terms as the $A$, $E$ and $T$ channels are approximately signal
orthogonal. In the low frequency limit the $A$ and $E$ can be shown~\cite{Crowder_Cornish} to be equivalent
to Cutler's $s_I$ and $s_{II}$~\cite{Cutler98} variables, which describe two Michelson detectors 
rotated by an angle ${\pi}/{4}$. Combining the geometrical factors $R_{12}$ with the noise transfer functions
leads to the effective sensitivity curves shown in Figures 7 and 8.

\begin{figure}[ht]
\includegraphics[angle=270,width=0.7\textwidth]{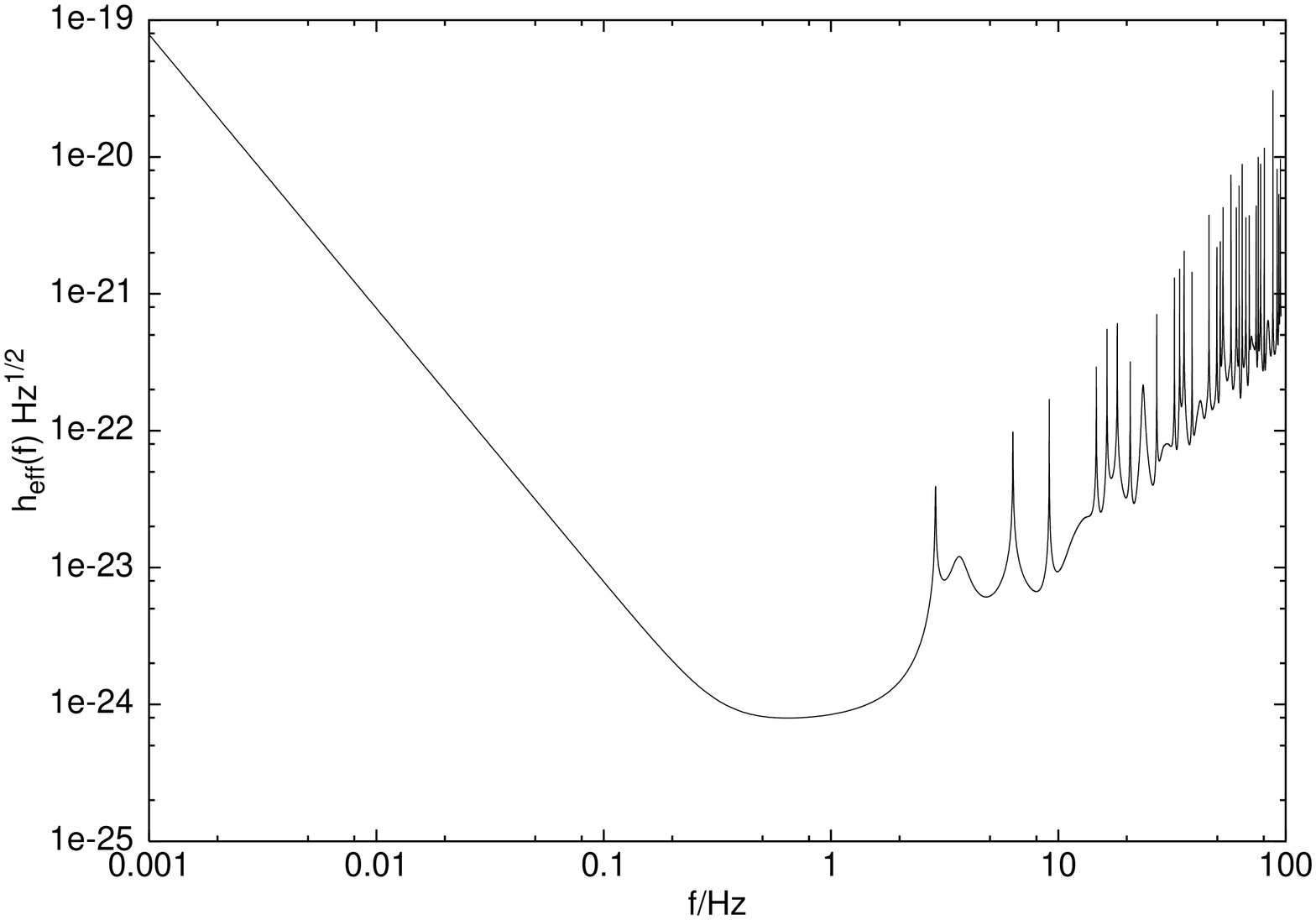}
\caption{Effective sensitivity for $A_{1} \times A_{2}$ ($E_{1} \times E_{2}$)}
\end{figure}
\begin{figure}[ht]
\includegraphics[angle=270,width=0.7\textwidth]{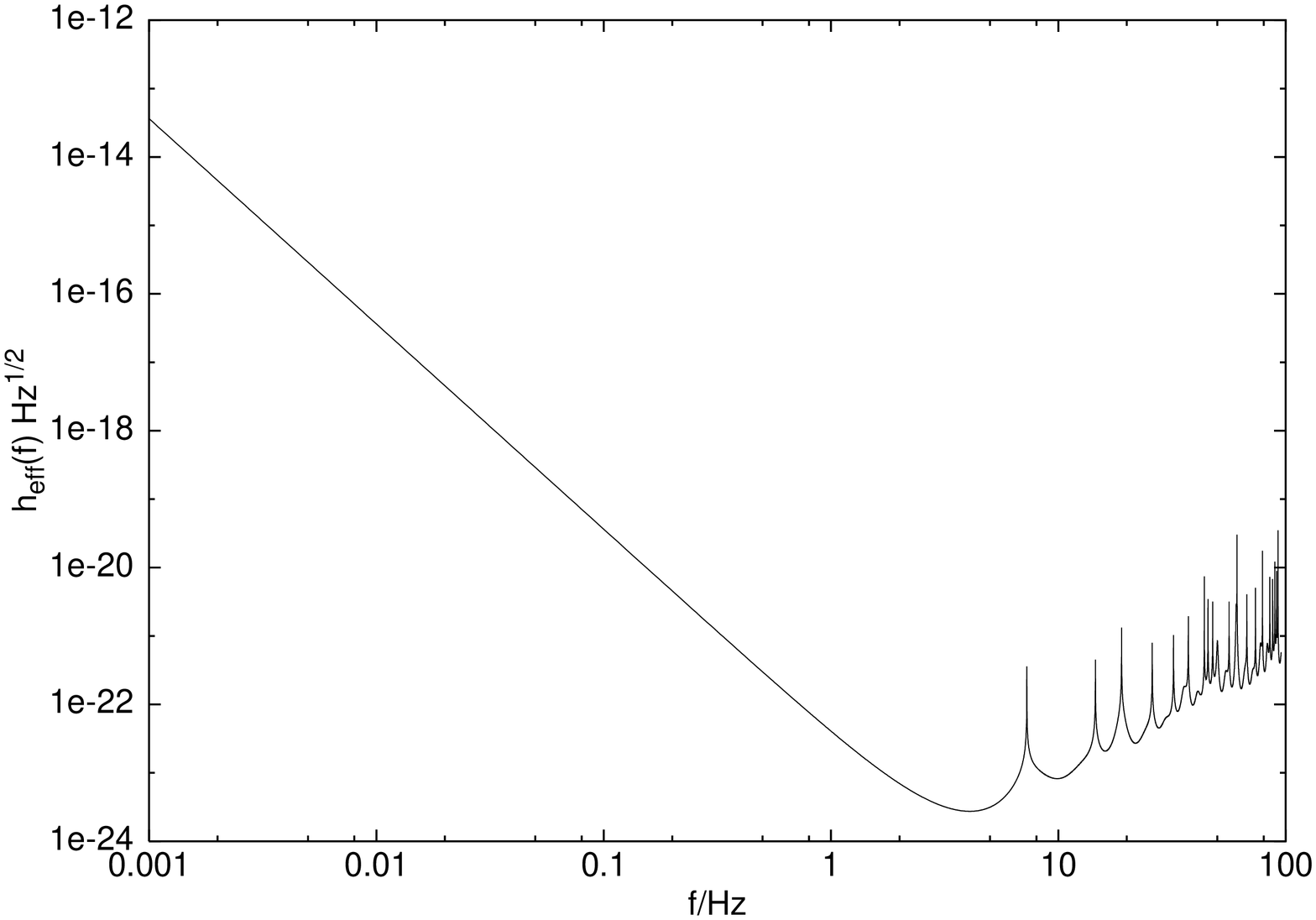}
\caption{Effective sensitivity for $T_{1} \times T_{2}$}
\end{figure}

The combined effective sensitivity follows from the optimal signal to noise ratio (\ref{opt_combo}) . 
Figure 9 shows the combined sensitivity curve using all channel combinations plotted against the
optimal CGB signal for a scale invariant spectrum with $\Omega_{gw} = 10^{-15}$.

\begin{figure}[ht]
\includegraphics[angle=270,width=0.7\textwidth]{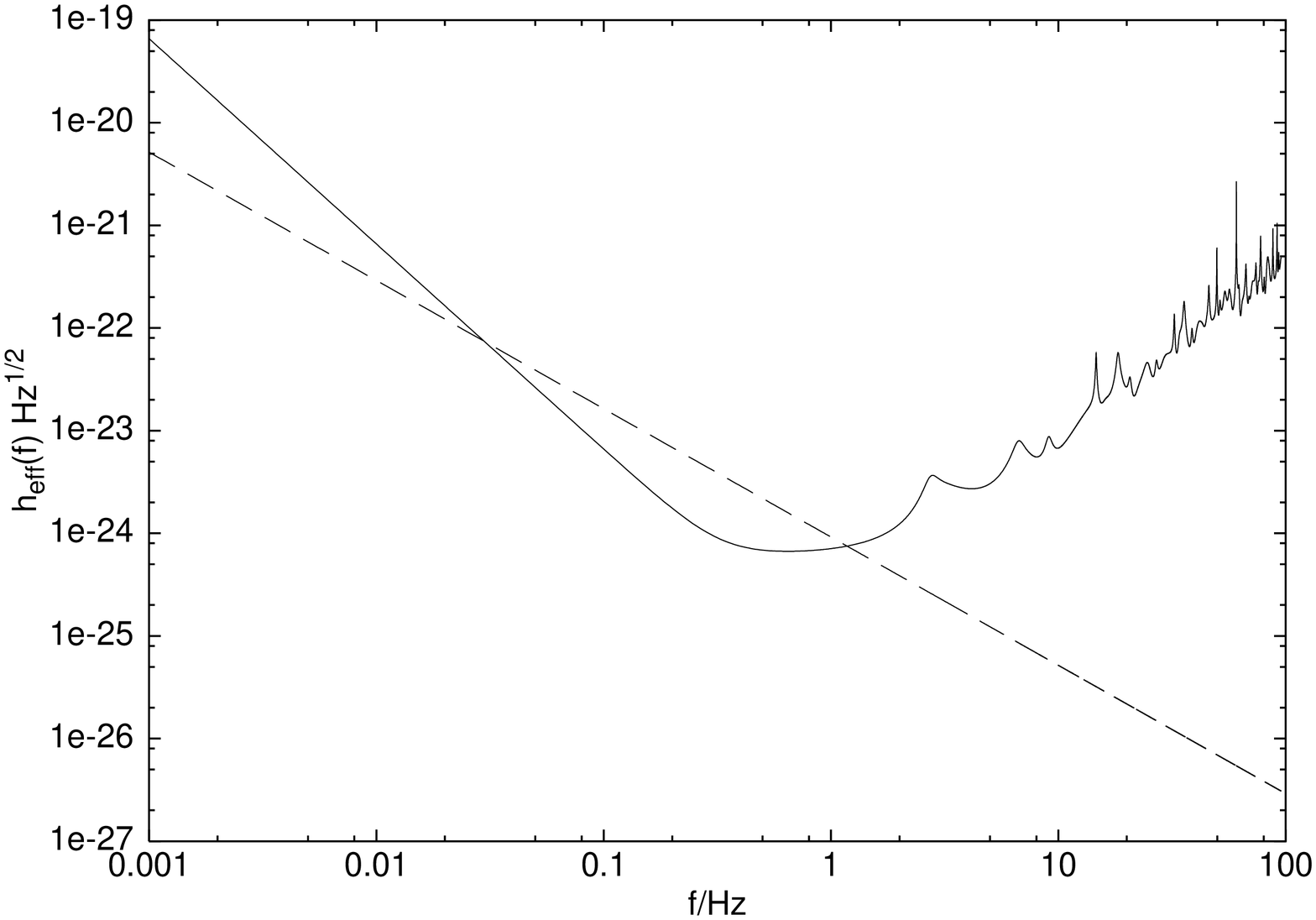}
\caption{The combined sensitivity curve
(solid line) plotted against the optimal CGB signal (dashed line) for an observation time of $T=1 {\rm yr}$.}
\end{figure}

An alternative way of conveying the information contained in Figure 9 is to plot 
${\rm SNR}(f)=(d {\rm SNR}^2(f)/ d \ln f)^{1/2}$: the contribution to the signal
to noise ratio per logarithmic frequency interval. The optimal ${\rm SNR}(f)$ for
$\Omega_{gw}=10^{-15}$ is shown in Figure 10.

\begin{figure}[h]
\includegraphics[angle=270,width=0.7\textwidth]{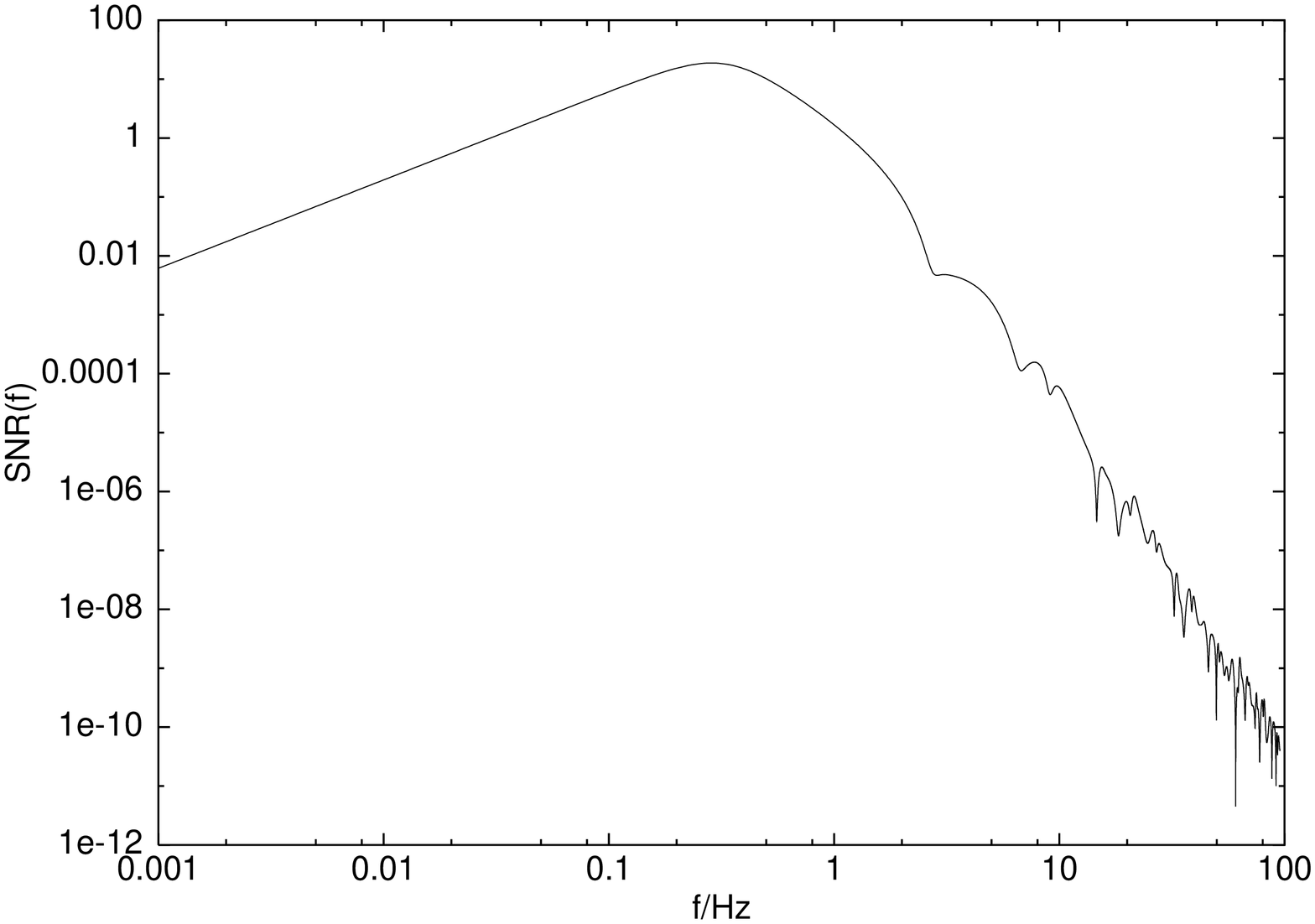}
\caption{Optimal ${\rm SNR}(f)$}
\end{figure}

The preceding graphs assumed that the two triangular interferometers
formed a symmetrical six pointed star. However, one could consider alternative
arrangements where the two detectors are rotated by an arbitrary angle
$\lambda$ with respect to one another. In our numbering convention for the
vertices of each triangle the symmetric star corresponds to a rotation
angle of $\lambda = \pi$. As we vary $\lambda$, the individual SNRs from $A_{1} \times A_{2}$
and $E_{1} \times E_{2}$ decrease, with minima at $\lambda = {\pi}/{4}$
and $\lambda = {3\pi}/{4}$ as expected~\cite{njc01}. The SNR for $A_{1} \times E_{2}$,
on the other hand, increases, with maxima at $\lambda = {\pi}/{4}$ and
$\lambda = {3\pi}/{4}$. This behavior is illustrated in Figures 11 and 12.
The net effect is that the optimal SNR is independent
of the relative orientation $\lambda$.

\begin{figure}[h]
\includegraphics[angle=270,width=0.7\textwidth]{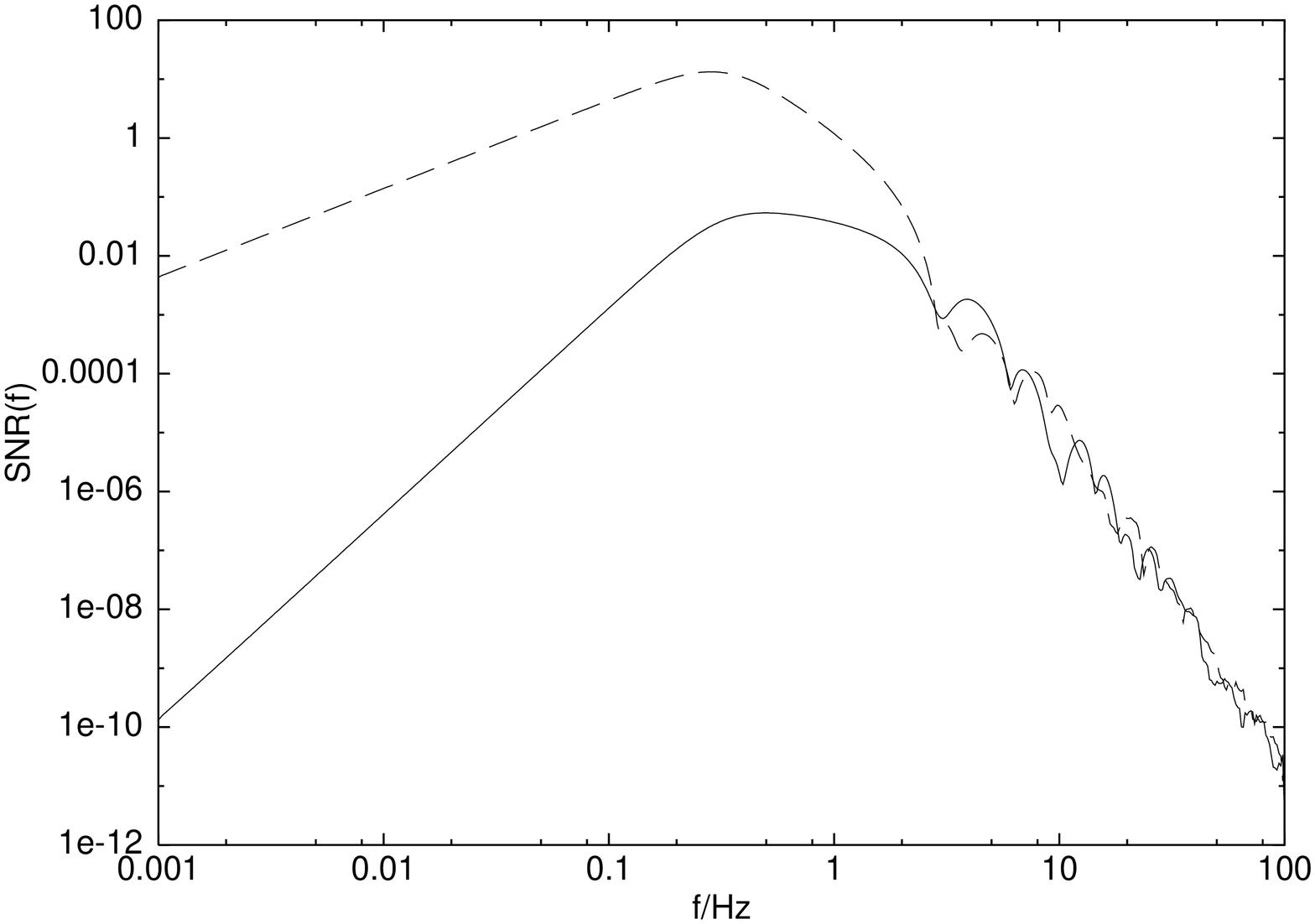}
\caption{${\rm SNR}_{A_{1}A_{2}}(f)$ for an angle of $\lambda=\pi$ (dashed line) and $\lambda=\pi/4$ (solid line).}
\end{figure}

\begin{figure}[h]
\includegraphics[angle=270,width=0.7\textwidth]{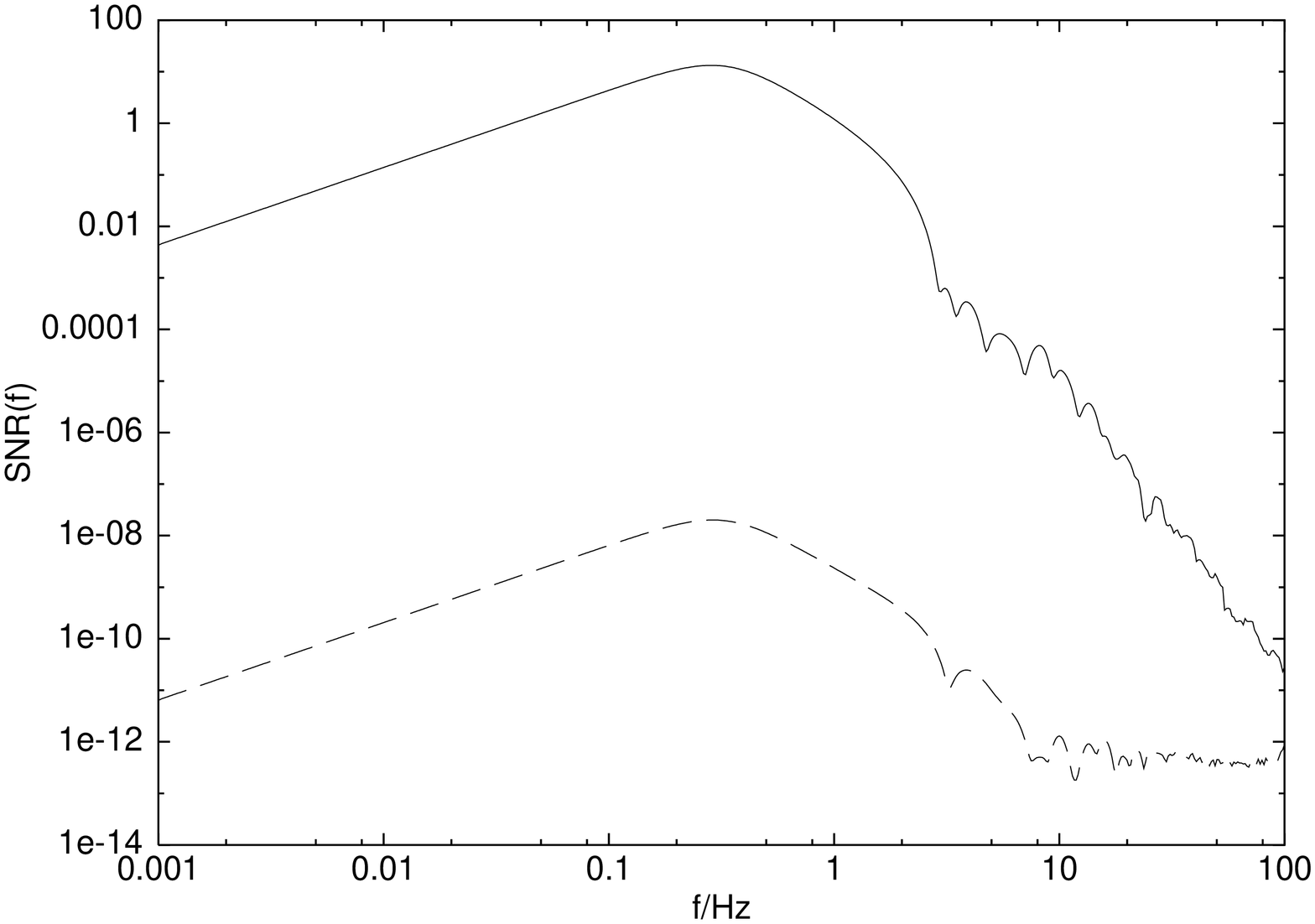}
\caption{${\rm SNR}_{A_{1}E_{2}}(f)$ for an angle of $\lambda=\pi$ (dashed line) and $\lambda=\pi/4$ (solid line).}
\end{figure}

We find that the optimal SNR for scale invariant CGB that can be achieved by the fiducial BBO design is equal to
\begin{equation}
{\rm SNR}_{\rm opt} = 155\sqrt{\frac{T}{5 {\rm yr}}}\frac{\Omega_{gw}}{10^{-15}}
\left(\frac{H_{0}}{70 \, {\rm km \,s}^{-1} \, {\rm Mpc}^{-1}}\right)^{2} \, ,
\end{equation}
which is well above the 3.3 threshold mentioned in the introduction. Conversely, the minimum $\Omega_{gw}$ for which we could detect
the CGB with the 95\% confidence is equal to $2.2 \times 10^{-17}({5 {\rm yr}}/{T})^{1/2}(70\, {\rm km \,s}^{-1} \, {\rm Mpc}^{-1}/{H_{0}})^2$.
Our findings agree with the recent independent calculation by Seto~\cite{seto05}.

\section{Conclusion}

We have determined that the fiducial BBO design would be able to detect a scale invariant CGB with an energy density as low
as $\Omega_{gw} = 2.2 \times 10^{-17}({5 {\rm yr}}/{T})^{1/2}(70\, {\rm km \, s}^{-1} \, {\rm Mpc}^{-1}/{H_{0}})^2$. We found
that the optimal sensitivity is independent of the relative orientation of the co-planar detectors used to perform the
cross correlation.

\end{document}